\documentclass[a4paper,twoside]{article}

\usepackage{epsfig}
\usepackage{subcaption}
\usepackage{calc}
\usepackage{amssymb}
\usepackage{amstext}
\usepackage{amsmath}
\usepackage{amsthm}
\newtheorem{proposition}{Proposition}[section]   %

\usepackage[dvipsnames]{xcolor}  

\usepackage{multicol}
\usepackage{pslatex}
\usepackage{apalike}
\usepackage{SCITEPRESS}     
\usepackage{comment}
\usepackage{multirow}
\usepackage{adjustbox}

\usepackage{algorithm}
\usepackage[noend]{algpseudocode}

\begin{document}

\title{Extracting Body Text from Academic PDF Documents for Text Mining}

\author{\authorname{Changfeng Yu, Cheng Zhang and Jie Wang}
\affiliation{Department of Computer Science, University of Massachusetts, Lowell, MA, U.S.A.}
\email{\{changfeng\_yu, cheng\_zhang\}@student.uml.edu, wang@cs.uml.edu}
}

\keywords{body-text extraction, HTML replication of PDF, line sweeping, backward traversal}

\abstract{
Accurate extraction of body text from PDF-formatted academic documents is essential 
in text-mining applications for deeper semantic understandings.
The objective is to extract complete sentences
 in the body text into a txt file with the original sentence flow and paragraph boundaries.
Existing tools for extracting text from
PDF documents 
would often mix body and nonbody texts.
We devise and implement a system called PDFBoT to
detect multiple-column layouts using a line-sweeping technique,
remove nonbody text using
computed text features and syntactic tagging in backward traversal, and
align the remaining text back to sentences and paragraphs.
We show that PDFBoT is highly accurate with average F1 scores of,
respectively, 0.99 on extracting sentences, 0.96 on extracting paragraphs,
and 0.98 on removing text on tables, figures, and charts
 over a corpus of 
PDF documents randomly selected from arXiv.org
across multiple academic disciplines.}

\onecolumn \maketitle \normalsize \setcounter{footnote}{0} \vfill

\section{\uppercase{Introduction}}
\label{sec:1}

\noindent It is desirable for text mining applications to extract complete sentences and correct boundaries of paragraphs 
from the body text of a PDF document into a txt file without hard breaks inside 
each paragraph.
Layered reading (http://dooyeed.com) and extractive summarization, for example,
are such applications. Layered reading
allows the reader to read the most important layer of sentences first based on sentence rankings, then the layer of next important sentences interleaving with the previous layers of sentences in the original order of the document, and continue in this fashion until the entire document is read. 

By ``body text" (BT in short) it means the main text of 
an article, 
excluding ``nonbondy text" (NBT in short) such as
headings, footings, sidings (i.e., text on side margins),
tables, figures, charts, captions, titles, authors, affiliations, and
math expressions in the display mode, among other things. 
%

%
%
%
Most existing tools for extracting text from PDF documents, including pdftotext (FooLabs, 2014) 
and PDFBox (Apache, 2017), 
extract a mixture of both BT and NBT texts. 
Identifying BT text from such mixtures of texts is challenging, if not impossible.
Other tools extract
texts according to rhetorical categories such as LA-PDFText (Burns, 2013) 
and logical text blocks such as
Icecite (Korzen, 2017), 
which only provide a suboptimal solution to our applications.

Extracting BT text from PDF documents of arbitary layouts is challenging, due to the utmost flexibility of PDF typesetting. 
Instead, we focus on BT extraction from 
single-column and multiple column research papers, reports, and case studies.
We do so by working with the location, font size, and font style of each character, and the locations and sizes of other objects. While a PDF file provides such information, we find it
easier to work with HTML replications produced by an exiting tool named pdf2htmlEX (Wang, 2014),
with almost the same look and feel of the original PDF document,
providing necessary formatting information via HTML tags, classes, and id's in the underlying DOM tree. 

We devise a system named PDFBoT (PDF to Body Text)
that, using pdf2htmlEX as a black box,
incorporates certain text formatting features produced by it to identify NBT texts. 
We use a line-sweeping method
to detect multi-column layouts and  
the area for printing the BT text. We also
develop multiple tests to identify NBT text inside the BT-text area
and use a backward traversal method to deploy these tests.
In addition, we use POS (Part-of-Speech) tagging 
to help identify NBT text that  are harder to
distinguish. 

%
The rest of the paper is organized as follows:
Section \ref{sec:2} is related work on text extractions from PDF.
Section \ref{sec:3} describes HTML replications via pdf2htmlEX 
and
Sections \ref{sec:4} presents the architecture of PDFBoT and its features
Section \ref{sec:6} is evaluation results with F1 scores and running time, and
Section \ref{sec:8} is conclusions and final remarks.

\section{\uppercase{Related Work}}
\label{sec:2}

\noindent 
Existing tools, such as
pdftotext (FooLabs, 2014) 
and
PDFBox (Apache, 2017) 
the two most widely-used tools for extracting text from PDF,
and a number of other tools such as
pdftohtml (Kruk, 2013), 
pdftoxml (Dejean and Giguet, 2016), 
pdf2xml  (Tiedemann, 2016), 
ParsCit (Kan, 2016), 
PDFMiner (Shinyama, 2016), 
pdfXtk (Hassan, 2013), 
pdf-extract (Ward, 2015), 
pdfx (Constantin et al, 2011), 
PDFExtract (Berg, 2011), 
and Grobid (Lopez, 2017), 
extract text from PDF extract BT text and NBT text together without
a clear distinction.
PDFBox can extract text in two-column layouts; some other 
tools  extract text line by line across columns.

Using heuristics is a common approach.
For example, the Java PDF library
was used to
obtain a bounding box for each word, compute the distance between neighboring 
words, connect them based on a set of rules
to form a larger text block, place them into rhetorical categories, and 
connect these categories following the order of the underlying document  \cite{Ramakrishnan2012}.
However, this method fails to align broken sentence and determine
text on formulas, tables, or figures. 
Using an intermediate HTML representation generated by pdftohtml 
\cite{yildiz2005}.
Text blocks may also be created by grouping characters based on their relative positions \cite{tableRec2016}, while extracting the tables in PDF.  
These two methods are focused only on extracting tables.

Other methods include rule-based and machine-learning models. 
For example, 
text may be placed into predefined logical text blocks
based on a set of rules on the distance, positions, fonts of characters, words, and text lines \cite{Bast2017}. 
However, these rules also connect text on tables or figures as BT text.
A Conditional Random Field (CRF) model is trained \cite{Luong2011,Romary2015} 
to extract texts 
according to a predefined
rhetorical category, such as title, abstract, and
other sections in the input document. 
However, this model fails to determine paragraph boundaries or
align broken sentences, among other things.

CiteSeerX \cite{citeseer}, a search engine, extracts metadata from indexed articles in scientific documents for searching purpose, but not focused on the accuracy of extracting body text. 
PDFfigures \cite{pdffigures} chunks the text table and figure into blocks, then classifies these blocks into captions, body text, and part-of-figure text.
Recent studies have shifted attentions to extracting certain types of text,
including titles \cite{Yang2019} (but not text on tables or figures), and
math expressions in the display mode and the inline mode \cite{Mali2020,Pfahler2019,Wang2018,Phong2020}.
%

In summary, 
previous methods, while meeting with
certain success, 
still fall short of the
desired accuracy required by text-mining applications
relying on clean extractions of complete sentences and correct
boundaries of paragraphs in BT text.

\section{\uppercase{HTML Replication of PDF}}
\label{sec:3}

\noindent HTML technologies have been used to replicate PDF layouts to facilitate online publishing.
A PDF document can be represented as a sequence of pages, with
each page being a DOM tree of objects with sufficient information 
for an HTML viewer to display the content \cite{Wang2013}. 
%
The text extracted from PDF by pdf2htmlEX (Wang, 2014) 
are translated into HTML text elements that are placed into the same positions as they are displayed by PDF. 

Let $F$ denote a PDF document and $f$ the HTML file produced by pdf2htmlEX on $F$.
The DOM tree for $f$, denoted by $T_f$, 
is divided into four levels: document, page,
text line, and text block (TBK in short). 

\textsl{(1) Document structure.}
$T_f$ starts with the following tag as the root: $\langle$div id=``page-container"$\rangle$, 
and each of its children is the root of a subtree for a page,
listed in sequence, with an id indicating its page number and 
a class name indicating the width and height of a page. For example, a child node with
$\langle$div id=``pf7" class=``pf w0 h0 data-page-no=``7"$\rangle$
is the root of the subtree for Page 7, where w0 and h0 are the width and height
of the page (specifying the printable area) with the origin at the lower-left corner of
the page.

\begin{figure}[h]
	\centering
	\includegraphics[width=0.5\columnwidth]{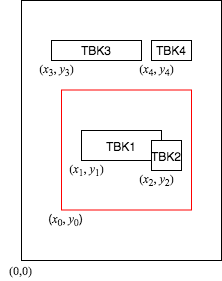}
	\caption{Schematic of the page structure. The
red square is a figure with a TBK1 and subscript TBK2 inside the square, where
$(x_0,y_0), (x_3, y_3), (x_4, y_4)$ are absolute coordinates, 
and $(x_i,y_i)$ ($1\leq i \leq 2$) are relative to $(x_0,y_0)$.
Thus, the corresponding absolute locations, denoted by $(x'_i, y'_i)$, are
$x'_i = x_0+x_i$ and $y'_i = y_0+y_i$ for $i=1,2$.}
	 \label{fig:htmlb.b}
\end{figure}

\textsl{(2) Page structure.} Each page starts with a page node, 
followed by object nodes with contents to be printed. 
Each object occupies 
a rectangular area (a bounding box) specified on a coordinate system of pixels.
The text of the document is divided into TBKs as
leaf nodes. 
Each TBK is represented by a $\langle$div$\rangle$ tag with corresponding attributes, and so the text in a TBK are either all BT text or all NBT text. 
%
%
Each object is identified by coordinates $(x,y)$ at the lower-left
corner of the bounding box relative to the coordinates of its parent node.
In what follows, these coordinates are referred to as the \textsl{starting point}
of the underlying object.
In addition to the starting point, a non-textual object is specified by a width and a height, and a TBK is specified with a height without
a width, where 
the width is implied by the enclosed text, font size and style, and word spacing.
The parent of each object may either be the origin, a node for a figure or a table,
or a node due to some (probably invisible) formatting code. 
Thus, the height of a page's DOM tree could be 
greater than 3. 
Figure \ref{fig:htmlb.b} is a schematic of page structure. 

\textsl{(3) Line structure.}
Each horizontal text line is made up of one or more TBKs, 
and no horizontal TBK  contains text across multiple lines. 
But TBKs in the NBT text could span across multiple lines, which are either
vertical or diagonal, specified by webkit-transform rotations, 
which rotates the text box around the center of the text box.
For example, the background text of ``Unpublished working draft"
and ``Not for distribution" on certain documents are two diagonal TBKs
on top of BT text.

\textsl{(4) Text-block structure. }
Each TBK is specified by exactly eleven classes of features,
where each feature class consists of one or more features, including
starting point $(x,y)$ relative to the starting point of its parent, 
height, font size, font style, font color, and word spacing.
Enclosed in TBK are text and additional spacing between words.
A TBK ends either at the end of a line
or at the beginning of a subscript, a superscript, and a citation.

%

\section{\uppercase{PDFB\lowercase{o}T}}
\label{sec:4}

\begin{figure}[h]
	\centering
	\includegraphics[width=0.9\columnwidth]{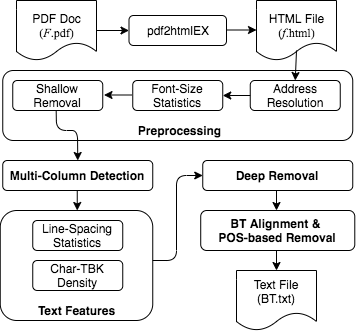}
	\caption{PDFBoT architecture and data flow diagram} 
	\label{fig:PDFBoT}
\end{figure}

\noindent 
PDFBoT consists of five major components: Preprocessing, Multi-Column Detection, Text Features, 
Deep Removal, and BT Alignment \& POS-based Removal. Figure \ref{fig:PDFBoT}
depicts the architecture and data flow diagram of PDFBoT.

\subsection{Preprocessing}


\textsl{(1) Address resolution.}
On each page in the DOM tree $T_f$,
each object occupies a rectangular area, specified by the starting point
relative to the starting point of its parent node, and some other formatting features.
The Preprocessing component calculates the absolute starting point of each object by a breadth-first search of the DOM tree. 
The starting points of the objects at the first level are already absolute.
For each object at the second level or below, 
let $(x,y)$ be its
relative starting point and $(x'_p,y'_p)$ the absolute starting point of its parent.
Then its absolute starting point is determined by $(x',y') = (x+x'_p, y+y'_p).$
In what follows, when we mention a starting point of a TBK,
we will mean its absolute starting point, unless otherwise stated.

\textsl{(2) Font-size statistics.}
This module  computes the frequency of each font size (over the total number of characters)
by traversing each TBK to obtain its font size and the number of characters in the text it encloses.
The font size with the highest frequency, denoted by BASE\_FS, is the font size for BT.

\textsl{(3) Shallow removal.}
This module removes
all non-textual objects (images and lines) and all TBKs  with font size 
beyond the interval 
$${\mathcal I}_f = (\mbox{BASE\_FS}- \Delta_2, \mbox{BASE\_FS}+ \Delta_2),$$
where $\Delta_2$ is a threshold value (e.g., $\Delta_2 = 3$),
or with a rotated display, which can be checked by its webkit-transform matrix.
Headings, sidings, and footings tend to have smaller font sizes than $\text{BASE\_FS}-\Delta_2$ (except page numbers)
and so they are removed by this module.

\textsl{Remark.} The abstract 
may have a slightly smaller font size
than $\mbox{BASE\_FS}$ (such as 3 pt smaller as in this paper). Setting an appropriate
value of $\Delta_2$ can resolve this problem. We may also deal with the abstract
separately, regardless its font size, using the keyword ``Abstract" and
the keyword ``Introduction" to extract the abstract.


\subsection{Multi-Column Detection}

Most lines on a given column are aligned flush left, except that the first line in a paragraph may be indented.
Start a vertical line sweep on each page from the left edge to the
right-hand edge one pixel at a time.
Let $n_p(i)$ denote the number of $x$-coordinates in the starting points 
of TBKs that are equal to $i$ on page $p$, where
$i$ starts from 0 and ends at $W$ one pixel at a time, and $W$ is the width of the printable area of the page (typically just the width of the page).
Note that a TBK does not have coordinates at the right-hand side. 

A line is aligned flush left to a column if 
the $x$-coordinate of the starting point of the leftmost TBK in the line
is equal to the $x$-coordinate of the left boundary of the said column.
It is reasonable to assume that (1) the left boundary of a corresponding column
is at the same $x$-coordinate on all pages and
(2) over one-half of the lines in any column across all pages 
are aligned flush left on each page. We also assume the following:
Let $j$ be the left boundary of a column. If $i$ is not the left boundary of a column, then
 $\sum_p n_p(i)$ (summing up $n_p(i)$ over all pages) is substantially
smaller than $\sum_p n_p(j)$.

\begin{proposition} \label{prop:2}
A document has $k$ columns ($k \geq 1$) iff the function
$\sum_p n_p(i)$ 
has exactly $k$ peaks with about the same values, and
the $i$-th $x$-coordinate that registers a peak is the left boundary of the $i$-th column.
\end{proposition}

\textsl{Remarks.} (1) Columns may begin at different $x$-coordi\-nates 
for pages that are even or odd numbered. 
Just treat pages of even (and odd) numbered as one document and
then Proposition \ref{prop:2} applies to them respectively.
(2) A two-column layout may have a one-column layout inserted, such as
a one-column abstract in a two-column academic paper. 
This can be detected
by checking the locations of TBKs. If most of them do not match with
the $x$-coordinate for the second column, then the underlying portion of the text
is a single column. Single-column text is processed in the same way as the left-column text.
(3) A more sophisticated method is to use a shorter vertical line  segment
to cover a sufficient number of lines for sweeping each time,
and move this line segment as a vertical sliding window.

\subsection{Text Features}
 
\textsl{(1) Line-spacing statistics.}
This module lines up TBKs according to their starting points to form lines in sequence.
Let $(x_1,y_1)$ and $(x_2,y_2)$ be the starting points of two text blocks $B_1$ and $B_2$, respectively.
Then $B_1$ and $B_2$ are on the same line iff $|y_1 - y_2| \leq \Delta_1$ for a small fixed
value of $\Delta_1$. The purpose of allowing a small variation is to make typesetting 
more flexible to adjust and beautify the overall layout 
(e.g., $\Delta_1 = 5$). 
Suppose that they are on the same line, then
$B_1$ is at the left-side of $B_2$ iff $x_1 < x_2$. 
If they are not on the same line, then $B_1$ is on a line above that of $B_2$ iff 
$y_1 - y_2 > \Delta_1$. 
This gives rise to a Page-Line-TBK tree structure of depth 2,
where the Page node has Lines as children, and each Line node has
one or more TBKs as children.

The module then computes the gap between every two consecutive lines in each column 
and obtains the frequency for each gap. The most common gap is the line spacing in the body text, denoted by BASE\_LS.

\textsl{(2) Char-TBK density.}
This module computes, for each line $L$, the number of non-whitespace characters
over the number of TBKs contained in $L$. Denote by
 $\text{\#Char}_L$ and 
$\text{\#TBK}_L$, respectively, the number of non-whitespace characters and the number of TBKs contained in $L$. 
Define by $D_L$ the following density:
$D_L = \text{\#Char}_L/\text{\#TBK}_L.$
Let $\text{BASE\_CBD}$ denote the average Char-TBK density for the entire document.


\subsection{Deep Removal} \label{sec:deep}

This module removes NBT text with font sizes
within the range of ${\mathcal I}_f$. 
It is reasonable to assume the following features on a PDF document adhering
to conventional formatting styles:
(1) Math expressions in the display mode, text on tables, text of figures, text on charts, authors, and 
affiliations are indented by at least a pixel from the left boundary of the underlying column.
(2) Every sentence ends with a punctuation. If a sentence ends with a math expression
in the display mode, then the last line of the math expression must end with a punctuation. 
(3)
The first line of text followed
a standalone title is aligned flush left.

\textsl{(1) Remove sidings.} 
The BT area on each page is a rectangular area within which the BT text are printed. 
Depending on how the majority of the BT text are displayed, the underlying document
is of either single column or multiple columns.
A column for printing the BT text is referred to as a \textsl{major column}. 
A column on a side margin (such as the line numbers on some documents) 
is referred to as a \textsl{minor column}, where TBKs are in red boxes.
It is reasonable to assume that the width of a major column cannot be smaller than a 
certain value $\Gamma_1$ (e.g., $\Gamma_1 = 1.5$ inch = 144 pixels).
It is reasonable to assume that side margins are symmetrical. Namely, in the printable area, 
the width of the
left margin is the same as that of the right-hand margin. Without loss of generality,
assume that the width of a side margin is less than $\Gamma_1$.  
Most documents have either one major-column or two major-columns.
For a magazine layout, three major-columns may also be used. 
For example, the layout of this submission is of two columns.
%


\begin{proposition} \label{prop:3}
Let $k$ be the number of columns (as detected by line sweeping as in Proposition \ref{prop:2}).
Let $w_m$ denote the width of a side margin.
Initially, set $w_m \leftarrow x_{b}$, the $x$-coordinate of the left boundary of the 
first column.
If $k>1$, let $x'_b$ be the $x$-coordinate of the left boundary of the second column.
If $x'_b - x_b < \Gamma_1$, then set $w_m \leftarrow x_b'$.
The BT area is from $w_m$ to $W-w_m$,
where 
$W$ is the width of the printable area of the page.
\end{proposition}
Note that if $k>1$ and $x'_b-x_b < \Gamma_1$, then the first column is not a major column. 
 Any TBKs with an $x$-coordinate of its starting point less
than $w_m$ is on the left margin and any TBKs with an $x$-coordinate of its starting point 
greater than $W - w_m $ is on the right-hand margin.
For example, this method removes line numbers. 
On a different formatting
we have encountered, such as on the \LaTeX template
for submitting drafts to a journal by the IOS Press, 
a line number
is a TBK with a starting $x$-coordinate in the left margin, where
the text enclosed is a pair of the same number with a long whitespace inserted in between that 
crosses over the entire BT text from left to right. This pair of numbers will also be removed
because its starting point is in the left margin.

\textsl{(2) Remove references.}
%
%
The simplest way to detect references is to search for
a line that consists of only one word ``References" that is
either on the first line of a column or has a larger space than 
BASE\_LS.
%
%
Remove 
everything after (this may remove appendices, which for our purpose is acceptable).
A more sophisticated method is to use the following line-sweeping method to detect the area of references. 
by detecting nested columns within a major column 
and Proposition \ref{prop:3}): Start from one pixel after the left boundary of a major column,
sweep the column from left to right with a vertical line on the entire paper.
If a local peak occurs with the same $x$-coordinate on consecutive lines,
each line from the left boundary of the column to this $x$-coordinate
is either null or a numbering TBK. A numbering TBK contains a number inside.
Then any line that has this property is a reference.
To improve detection accuracy, we may also use
 a named-entity tagger \cite{peters2017semi} to determine if the text right after
 a numbering TBK are tagged as person(s).

\textsl{(3) Remove special lines.}
Let $x_c$ be the $x$-coordi\-nate of the left boundary of the column that line $L$ belongs to,
 and $x_L$ be the $x$-coordinate in the starting point of the leftmost TBK in $L$.
If $x_L-x_c > \Gamma_2$ for a fixed value of $\Gamma_2$ larger
than normal indentation (e.g., $\Gamma_2 = 50$; normal indentation for a paragraph is 48 pixels or less), then remove $L$.
%
This module removes most of the math expressions in the display mode,
certain author names and affiliations, as well as
text on figures 
with the same font size as the BT text, for in this case the leftmost TBK would
have a large indentation due to the space taken by the $y$-axis and a vertical title.


If line $L$ contains a TBK that includes a whitespace greater than a certain threshold $\Gamma_3$ (e.g. $\Gamma_3 = 50$), specified by
a $\langle$span$\rangle$ tag, then remove $L$.
%
%
It is evident that such a TBK is NBT. 

\textsl{(4) Remove lines by backward scans and NBT tests.}
The following tests are used in certain combination to determine NBT text lines.

 \textsl{(a) Line-Spacing Test.} 
An NBT line typically has larger line spacing (gap) from the immediate line above and from the immediate line below. A line $L$ passes the line-spacing test If 
the gap from $L$ to the immediate line above (if it exists) and the immediate line below (if it exists)
is either too large or too small; namely, it is beyond
an interval 
$${\mathcal I}_{g} = (\text{BASE\_LS} - \Gamma_4, \text{BASE\_LS} + \Gamma_4)$$
for a certain threshold $\Gamma_4$
(e.g.,
$\Gamma_4 = \Delta_2 = 3$).  

\textsl{(b) Char-TBK Density Test.}
A line $L$ in math expression in the display mode 
typically consists of a larger number of short TBKs because of the presence of subscripts and superscripts, 
where each word or symbol would be
by itself a TBK. Thus, 
the char-TBK density $D_L$ would be much smaller than BASE\_CBD, 
the average Char-TBK density.
$L$ passes this test if
$D_L < \Gamma_5\cdot \text{BASE\_CBD}$ for a threshold value of $\Gamma_5$ (e.g., $\Gamma_5 = 10$). 

\textsl{(c) Punctuation Test.}
A line $L$ passes the punctuation test 
if the rightmost TBK in $L$ does not end with a punctuation. 

\textsl{(d) Indentation Test.}
A line $L$ passes the indentation test if the $x$-coordinate in the starting point of its
leftmost TBK is greater than that of the left boundary of the underlying column. 

\smallskip
\noindent
\textbf{NBT-Tests-based Removal Algorithm.} 
On a given document, 
scan text from the line preceding the list of references
and move backward one page at a time 
to the first line on the first page.
On each page, scan from the bottom line in the rightmost column and move up one line
at a time. Once it reaches the top line, scan from the bottom line in the column on
the left and move up one line at a time. When the top line on the leftmost column is reached,
move backward to the preceding page and repeat. 
Let $P$ be a Boolean variable.
%
Initially, set $P \leftarrow 0$.
Scan text lines in the aforementioned order of traversal.
In general, if a line is kept, then set $P$ to 0. 
If a line is removed, then set $P$ to 1, unless otherwise stated.

In particular, do the following during scanning:
(1) If $L$ passes both  of the indentation test and the char-TBK density test, then
remove $L$ and set $P \leftarrow 1$.
(2) If $P=0$ and $L$ passes the line-spacing test and the punctuation test,
then remove $L$ and set $P\leftarrow 0$.
(3) Otherwise, keep $L$ and set $P \leftarrow 0$.

Rule 1 removes page numbers, authors and affiliations,
text on tables, text on figures, and text on charts that pass both of
the indentation test and the char-TBK density test.
This rule also removes math expressions in the display mode.
It does not remove the last line of a paragraph because such a line fails the indentation
test.
It does not remove
a single-sentence paragraph as long as
it is not too short and does not contain multiple TBKs, 
for it would defy the small char-TBK density test.
It does not remove a text line that contains an inline math expression
as long as it is not the first line in an indented paragraph.

Rule 2 removes standalone one-line and two-line titles that are 
not ended with a punctuation in each line for the following reason: By assumption, 
the first text line below a standalone title is aligned flushed left
and so it will not be removed, which means that $P = 0$ (see Rule 3).
Likewise, this rule also removes captions without punctuation at the end
of each line, if its successor line is not removed, which implies that $P=0$ (see Item 3 below).
This rule does not remove the last line in a math expression in the display mode
for this line must end with a punctuation by assumption, which means that $P=1$.
This ensures that the line preceding the displayed math expression
that doesn't end with
a punctuation is not removed by this rule.

\subsection{BT Alignment \& Syntactic Removal}

After Deep Removal, PDFBoT aligns BT lines 
to restore sentences and paragraphs without hard breaks.
Recall that lines are formed according to columns. 
For each page, BT Alignment starts from the first line in the leftmost column one line at a time
and removes hard breaks within a paragraph until the last line in the current column. Then it moves to the next column (if there is any) and repeat the same procedure until the last line in the last column. In addition to removing hard breaks within a paragraph, it also needs to take
special care of hyphens at the end of a line and boundaries of paragraphs.
Removing hyphens at the end of lines is the easiest way. While this might 
break a hyphenated word into two words, doing so has a minor impact on our task while having a much larger benefit of restoring a word.
We may also use a dictionary to determine if a hyphen at the end of a line belongs to a
hyphenated word and keep it if it does.

If a line $L$ meets one of the following three conditions, then it is the first sentence of a paragraph: 
(1)
The gap between $L$ and the immediate line above is greater than $\text{BASE\_LS} + \Gamma_4$. 
(2) The $x$-coordinate of the leftmost TBK in $L$ is larger
than that of the leftmost TBK in the line immediately above.
The rest is text extraction from each TBK in the order of line locations.
Denote by $f'$ the txt file from this process.

%
While Shallow Removal and Deep Removal can remove most of the NBT-text lines,
captions that end with punctuation could still remain in BT text.
To remove all captions, we use the line-spacing rule to
group lines in a caption in $f'$ into a paragraph.
In this paragraph, the first keyword would
be one of the followings: ``Table", ``Figure", ``Fig.", followed by a string of
digits and dot. If the third word in the first line of such a paragraph
is not a verb, then this paragraph is deemed to be a caption.
We use an existing tool \cite{toutanova2003feature} to obtain  part-of-speech (POS) tags for each such paragraph, and remove it accordingly. 
Let BT.txt be the output.


Let $T_1(F)$ and $T_2(f')$ denote, respectively, the
time complexities of pdf2htmlEX on PDF file $F$ and POS tagging on 
paragraphs starting with ``Table", ``Figure", or ``Fig." in $f'$. 

\begin{proposition} \label{prop:4}
PDFBoT runs in $T_1(F)+T_2(f')+O(np)$ time on an input PDF document $F$, where
$n$ is the number of pixels in the printable area of a page and $p$
is the number of pages contained in $f$ generated by pdf2htmlEX.
\end{proposition}

\subsection{Display Sentences in Colors}

An optional 
component of PDFBoT, sentences may be colored in the original layout
of the HTML replicate by adding appropriate color tags in $f$.
%
%
Let ${\cal B} = \langle C_i\rangle_{i=1}^n$ represent the string of character objects of the BT text, where $C_i = (c_i, b_i, t_i)$ with $c_i$ being the $t_i$-th character in the text contained
in the $b_i$-th TBK.
Let $S = \langle l_j \rangle_{j=1}^{m}$ be the sentence to be highlighted,
where $l_i$ is the $i$-th character in $S$. 
Use a string-matching algorithm to find $\ell$ such that $\langle C_\ell, \ldots, C_{\ell+|S|-1} \rangle = S$. Let $startpoint = C_\ell$ and $endpoint = C_{\ell+|S|-1}$.

To color $S$ with a chosen color,
change the corresponding elements in $f$ as follows:
If $startpoint$ and $endingpoint$ are in the same TBK, 
add all the characters between $startpoint$ and $endingpoint$ to a new tag with 
an appropriate color attribute. 
Otherwise,
for the $startpoint$ block, add all the characters after $startpoint$ in the block to a new tag with a color attribute;
for the $endingpoint$ block, add all the characters before $endingpoint$ in the block to a new tag with the same color attribute;
and wrap all the TBKs between the $startpoint$ block and $endingpoint$ block with a new tag
with the same color attribute.


\noindent

\section{\uppercase{Evaluation}}
\label{sec:6}

\noindent 
We evaluate the accuracies of PDFBoT on the following tasks with a given document:
(1) Extracting complete sentences in the BT text.
(2) Getting correct boundaries of paragraphs.
(3) Removing text on tables and figures.
%
%
To do so,
we first need to determine an evaluation dataset.
To the best of our knowledge, no existing benchmarks are appropriate
 for evaluating PDFBoT. 
Bast and Korzen \cite{Bast2017}
presented a dataset on PDF articles collected from arXiv.org, where they
worked out a method to generate texts from the underlying \TeX or \LaTeX 
files as the ground-truth txt files for
evaluating extraction. However, this
dataset does not meet our need for the following reasons: 
(1) Most of the txt files do not contain Abstracts of the underlying PDF documents, and Abstracts are an important part of the BT text. (2) Some txt files contain authors and affiliations, and some don't, resulting in an inconsistency for evaluation. (3) The  txt files treat the text after
a math expression in the display mode as a new paragraph when it should not be.


We construct a dataset by selecting independently at random from arXiv.org 100 two-column PDF articles
in the disciplines of biology, computer science, finance, physics, and
mathematics with the following statistics on document sizes: (1) the average number of pages in an article: 8.28; (2) the median number of pages: 8; (3) the maximum number of pages: 17, and
the minimum number of pages: 4; the standard deviation: 2.94.
%
%
We manually compare the extracted text with the text in original academic PDF documents under three categories: sentences, paragraphs, and
text on tables and figures.

Possible outcomes 
for sentence and paragraph extractions
are (1) \textsl{correct}, (2) \textsl{erroneous}, and 
(3) \textsl{missing}, where ``correct" means
 that the sentences (paragraphs) extracted are BT text as the way they should be; 
``erroneous" for sentences means that either the sentence extracted is BT text but 
with an error, referred to as \textsl{incomplete}, or it should not be extracted at all, referred to as
\textsl{extra}, while ``erroneous" for paragraphs means that  the paragraph extracted is BT text but should not be a paragraph; 
and
``missing" means that a sentence (paragraph) should be extracted but
isn't.
%
%
%
Correct extraction is true positive (tp), erroneous extraction is false positive (fp), 
and extraction that is missing is false negative (fn).

Table \ref{table:sen1} is the statistics on extractions of
sentences and paragraphs, where \textsl{Total} means the total number of
true sentences and paragraphs, respectively, in the original articles.

\noindent
\begin{table}[h]
\centering
\caption{Statistics on extractions of sentences and paragraphs, where ``Incpl" means incomplete.} 
\label{table:par1}\label{table:sen1}
\begin{tabular}{c|c|c|c}
\hline
\multirow{2}{*}{\textbf{Total}} & \textbf{Correct} & \textbf{Erroneous} & \textbf{Missing}  \\
\cline{2-4} 
         & \textbf{(tp)}	& \textbf{(fp)}	& \textbf{(fn)} \\				
\hline
\multicolumn{4}{c}{Sentences}\\\hline
\multirow{2}{*}{19,564}  &  \multirow{2}{*}{19,158}   & 341 (Incpl)   	 &\multirow{2}{*}{30} \\
&&205 (Extra) & \\\hline
\multicolumn{4}{c}{Paragraphs}\\\hline
4,596 &  4,580	& 370 &19	\\
\hline
\end{tabular}
\end{table}

On removing text on tables, figures, and charts, 
possible outcomes are (1) \textsl{removed} and
(2) \textsl{remained}, where removed means
 that the text is correctly removed as it should be and
 remained means that the text that should be removed remains.
Removed is true positive and remained is false negative. Since every text
on a table or a figure/chart should be removed, 
there is no false positive. There are 9.469 TBKs on tables, figures, and charts in the corpus
with 8,986 TBKs correctly removed and 483 TBKs remained.

Table \ref{table:sen2} is the statistics of precision, recall, and F1 score,
which are computed individually and then rounded to the second decimal place, unless otherwise stated to avoid writing 1.00 due to rounding.

\begin{table}[h]
\centering
\caption{Sentence statistics of precision, recall, and F1 score.} 
\label{table:sen2}\label{table:par2}\label{table:tex2}
\begin{tabular}{l|c|c|c|c|c}
\hline
&\textbf{Avg} & \textbf{Med} & \textbf{Max} & \textbf{Min} & \textbf{Std} \\
\hline
\multicolumn{6}{c}{Sentences}\\\hline
Precision 	& 	0.97 	& 0.98 	& 1 	& 0.92 	& 0.02 \\
Recall 		&  0.999 	& 1 			& 1 	& 0.95  	& 0.01 \\   
F1 score 	& 	0.99 	& 0.99 	& 1	& 0.96	& 0.01 \\
\hline
\multicolumn{6}{c}{Paragraphs}\\\hline
Precision 	& 	0.93 	& 0.93 	& 1 			& 0.70	& 0.05 \\
Recall 		&  0.99 	& 1 			& 1 			& 0.81  	& 0.03 \\   
F1 score 	& 	0.96 	& 0.96 	& 1	& 0.83	& 0.03 \\
\hline
\multicolumn{6}{c}{Text on tables/figures/charts}\\\hline
Precision 	& 	0.93 	& 0.93 	& 1 			& 0.70	& 0.05 \\
Recall 		&  0.99 	& 1 			& 1 			& 0.81  	& 0.03 \\   
F1 score 	& 	0.96 	& 0.96 	& 1	& 0.83	& 0.03 \\
\hline
\end{tabular}
\end{table}

We note that in certain styles, paragraphs are not indented, but separated by 
an obvious line of whitespace. In this case, a text line that is not a new paragraph and
after a math expression in display mode could be mistakenly considered as a new paragraph.

Table \ref{table:time} is the running times incurred, respectively, by pdf2htmlEX and
PDFBoT after pdf2htmlEX generates a txt file on a 2015 commonplace laptop MacBook Pro with
a 2.7 GHz Dual-Core Intel Core i5 CPU and 8 GB RAM, 
where MAX represents the maximum running time in seconds processing a document in
this dataset, MIN the minimum running time, Avg the average running time,
Med the median running time, and Std the standard deviation.

\begin{table}[h]
\caption{Running time statistics (in seconds).}
\label{table:time}
\centering
\begin{tabular}{l|c|c|c|c|c}
  \hline
  &  \textbf{Avg} &  \textbf{Med} &  \textbf{Max} &  \textbf{Min} &  \textbf{ Std} \\
  \hline
  pdf2htmlEX & 3.00 & 1.90 & 13.8 & 0.80 & 2.47 \\
  PDFBoT  &  10.3 & 6.80 & 106 & 2.60& 12.5 \\   
  \hline
\end{tabular}
\end{table}

We note that the running time depends on how complex the content of the underlying document would be. 
It would take a substantially longer time to process if a document contains significantly 
more math expressions or tables.
A total of six documents each takes longer than 25 seconds
for PDFBoT to run. Checking these documents, we found that they
contain a large number of math expressions, tables, or supplemental materials after the references. The one extreme outlier that runs 106 seconds on PDFBoT but only 9.42 seconds on pdf2htmlEX is a 10-page
PDF document. The reason
 is likely due to complex features used to describe the
 document by pdf2htmlEX.
 While generating the HTML file would not be too costly,
 analyzing the CSS3 files to extract features for
 this particular document has taken more time, which
 needs to be investigated further.
 Overall, PDFBoT incurs 10.3 seconds on average.
 
\section{Conclusions and Final Remarks} 
\label{sec:8}
\noindent
PDFBoT 
uses certain formatting features, text-block statistics, syntactic features,
the line-sweeping method,
and the backward traversal method to
achieve accurate extraction. 
PDFBoT is available for public access at http://dooyeed.com:10080/pdfbot.

While the majority of the academic PDF documents satisfy the
assumptions listed in the paper, it is not always the case and so
some of the extraction mechanisms could fail. 
To further improve accuracy of detecting NBT text, particularly on a document
that violates some of the assumptions,
we may explore deeper features
in CSS3 files in addition to those we have used. 
For example, it would be useful to
investigate how to compute the width of a TBK.
Neural-network classifiers such as
CNN models may also be explored to identify certain types of NBT text  residing in the
BT text area.

\bibliographystyle{apalike}
{\small
\bibliography{KDIR_PDFBoT}}

\begin{thebibliography}{}

\bibitem[Bast and Korzen, 2017]{Bast2017}
Bast, H. and Korzen, C. (2017).
\newblock A benchmark and evaluation for text extraction from {PDF}.
\newblock In {\em ACM/IEEE Joint Conference on Digital Libraries (JCDL)}, pages
  1--10.

\bibitem[Clark and Divvala, 2015]{pdffigures}
Clark, C. and Divvala, S. (2015).
\newblock Looking beyond text: Extracting figures, tables, and captions from
  computer science paper.

\bibitem[Giles, 2006]{citeseer}
Giles, C.~L. (2006).
\newblock The future of citeseer: Citeseerx.
\newblock In F{\"u}rnkranz, J., Scheffer, T., and Spiliopoulou, M., editors,
  {\em Knowledge Discovery in Databases: PKDD 2006}, pages 2--2, Berlin,
  Heidelberg. Springer Berlin Heidelberg.

\bibitem[Luong et~al., 2011]{Luong2011}
Luong, M.-T., Nguyen, T.~D., and Kan, M.-Y. (2011).
\newblock Logical structure recovery in scholarly articles with rich document
  features.
\newblock {\em International Journal of Digital Library Systems (IJDLS)}, pages
  1--23.

\bibitem[Mali et~al., 2020]{Mali2020}
Mali, P., Kukkadapu, P., Mahdavi, M., and Zanibbi, R. (2020).
\newblock {\em ScanSSD: scanning single shot detector for mathematical formulas
  in PDF document images}.

\bibitem[Peters et~al., 2017]{peters2017semi}
Peters, M.~E., Ammar, W., Bhagavatula, C., and Power, R. (2017).
\newblock Semi-supervised sequence tagging with bidirectional language models.
\newblock {\em arXiv preprint arXiv:1705.00108}.

\bibitem[Pfahler et~al., 2019]{Pfahler2019}
Pfahler, L., Schill, J., and Mori, K. (2019).
\newblock The search for equations-learning to identify similarities between
  mathematical expressions.

\bibitem[Phong et~al., 2020]{Phong2020}
Phong, B.~H., Hoang, T.~M., and Le, T. (2020).
\newblock A hybrid method for mathematical expression detection in scientific
  document images.
\newblock {\em IEEE Access}, 8:83663--83684.

\bibitem[Ramakrishnan et~al., 2012]{Ramakrishnan2012}
Ramakrishnan, C., Patnia, A., Hovy, E., and Burns, G. (2012).
\newblock Layout-aware text extraction from full-text {PDF} of scientific
  articles.
\newblock {\em Source Code for Biology and Medicine}.

\bibitem[Romary and Lopez, 2015]{Romary2015}
Romary, L. and Lopez, P. (2015).
\newblock Grobid -- information extraction from scientific publications.
\newblock {\em ERCIM News}, 100. ffhal-01673305.

\bibitem[Shigarov et~al., 2016]{tableRec2016}
Shigarov, A., Mikhailov, A., and Altaev, A. (2016).
\newblock Configurable table structure recognition in untagged pdf documents.

\bibitem[Toutanova et~al., 2003]{toutanova2003feature}
Toutanova, K., Klein, D., Manning, C.~D., and Singer, Y. (2003).
\newblock Feature-rich part-of-speech tagging with a cyclic dependency network.
\newblock In {\em Proceedings of the 2003 conference of the North American
  chapter of the association for computational linguistics on human language
  technology-volume 1}, pages 173--180. Association for Computational
  Linguistics.

\bibitem[Wang and Liu, 2013]{Wang2013}
Wang, L. and Liu, W. (2013).
\newblock Online publishing via pdf2html{EX}.
\newblock {\em TUGboat}, 34:313--324.

\bibitem[Wang et~al., 2018]{Wang2018}
Wang, L., Wang, Y., Cai, D., Zhang, D., and Liu, X. (2018).
\newblock Translating math word problem to expression tree.
\newblock pages 1064–--1069.

\bibitem[Yang et~al., 2019]{Yang2019}
Yang, H., Aguirre, C.~A., Torre, M. F. D.~L., Christensen, D., Bobadilla, L.,
  Davich, E., Roth, J., Luo, L., Theis, Y., Lam, A., Han, T. Y.-J., Buttler,
  D., and Hsu, W.~H. (2019).
\newblock Pipelines for procedural information extraction from scientific
  literature: towards recipes using machine learning and data science.
\newblock pages 41--46.

\bibitem[Yildiz et~al., 2005]{yildiz2005}
Yildiz, B., Kaiser, K., and Miksch, S. (2005).
\newblock pdf2table: A method to extract table information from pdf files.
\newblock pages 1773--1785.

\end{thebibliography}

\end{document}